\documentclass[showpacs,10pt,preprintnumbers,amssymb,amsmath]{revtex4}
\usepackage[english]{babel}
\usepackage[matrix,arrow,curve]{xy}

\begin{document}

\bibliographystyle{apsrev}

\title{Isotropy Properties of the Multi-Step Markov Symbolic Sequences}

\author{S. S. Apostolov, Z. A. Mayzelis}
\affiliation{V. N. Karazin Kharkov National University, 4 Svoboda
Sq., Kharkov 61077, Ukraine}

\author{O. V. Usatenko\footnote[1]{usatenko@ire.kharkov.ua}, V. A. Yampol'skii}
\affiliation{A. Ya. Usikov Institute for Radiophysics and Electronics \\
Ukrainian Academy of Science, 12 Proskura Street, 61085 Kharkov,
Ukraine}

\begin{abstract}
A new object of the probability theory, the two-sided chain of
symbols (introduced in Ref.~arXiv:physics/0306170) is used to study
isotropy properties of binary multi-step Markov chains with the
long-range correlations. Established statistical correspondence
between the Markov chains and certain two-sided sequences allows us
to prove the isotropy properties of three classes of the Markov
chains. One of them is the important class of weakly correlated
additive Markov chains, which turned out to be equivalent to the
additive two-sided sequences.

\end{abstract}

\maketitle
%%%%%%%%%%%%%%%%%%%%%%%%%%%%%%%%%%%%%%%%%%%%%%%%%%%%%%%%%%%%
\section{Introduction}
\label{I} The problem of long-range correlated random symbolic
systems (LRCS) has been under study for a long time in many areas of
contemporary physics~\cite{bul,sok,bun,yan,maj,halvin},
biology~\cite{vossDNA,stan,buld,prov,yul,hao},
economics~\cite{stan,mant,zhang},
linguistics~\cite{schen,kant,kokol,ebeling,uyakm}, etc.

Among the ways to get a correct insight into the nature of
correlations of complex dynamic systems, the use of the
\emph{multi-step Markov} chains is one of the most important because
it allows constructing a random sequence with the prescribed
correlated properties in the most natural way. The $N$-step Markov
chains are characterized by the conditional probability that each
symbol of the sequence takes on some definite value depending on $N$
\textit{previous} symbols. These chains can be easily constructed by
sequential generation using the prescribed conditional probability
function. The binary correlation functions of the Markov chains can
be explicitly calculated in some simple cases. The concept of
additive chains turned out to be very useful because it is possibile
to evaluate the binary correlation function of the chain via the
memory function (see for the details
Refs.~\cite{uya,mel,AllMemStepCor}).

Another important reason for the study of Markov chains is their
application to the various physical objects~\cite{tsal,abe,den},
e.g., to the Ising chains of classical spins. The problem of
thermodynamical description of the Ising chains with long-range spin
interaction is open even for the 1D case. However, the association
of such systems with the Markov chains can shed light on the
non-extensive thermodynamics of the LRCS.

The LRCS can also be modeled by another class of correlated
sequences, the so-called two-sided chains introduced in
Sec.~\ref{def}. They are characterized by the conditional
probability that each symbol of the sequence takes on the definite
value depending on the neighboring symbols \emph{at both sides} of
the considered symbol. An example of systems with such a property is
the above-mentioned Ising chain. In Ref.~\cite{equiv}, we proved
that the two-sided sequences were statistically equivalent to the
Markov chains. In this paper this equivalence is used to study the
isotropy properties of the Markov chains.

The paper is organized as follows. In the first Section, we give the
definitions of the Markov and two-sided chains and formulate the
problem of this study. The next Section is devoted to the
examination of the anisotropy properties of three classes of Markov
chains: (a) the class of Markov chains that are equivalent to the
two-sided sequences with symmetric conditional probability function;
(b) the Markov chains with permutative conditional probability
function; (c) the additive Markov chains, which are shown to be
equivalent to the additive two-sided sequences.
%%%%%%%%%%%%%%%%%%%%%%%%%%%%%%%%%%%%%%%%%%%%%%%%%%%%%%%%%%%%
\section{General definitions}
\label{def}

Let us determine the \textit{$N$-step Markov} chain. This is a
sequence of random symbols,~$a_i$, $i\in
\mathbb{Z}=\{\dots,-2,-1,0,1,2,\dots\}$, possessing the following
property: the probability of symbol~$a_i$ to have a certain value,
under the condition that the values of all \textbf{previous} symbols
are fixed, depends on the values of~$N$ previous symbols only,
\begin{equation}\label{def_mark}
 P(a_i=a|\ldots,a_{i-2},a_{i-1})=
 P_N(a_i=a|a_{i-N},\ldots,a_{i-2},a_{i-1}).
\end{equation}
The Markov chain is a \textit{homogeneous} sequence, because the
conditional probability~Eq.~(\ref{def_mark}) does not depend
explicitly on $i$, i.e., is independent of the position of symbols
$a_{i-N},\ldots,a_{i-1},a_{i}$ in the chain. It only depends on the
values of $a_{i-N},\ldots,a_{i-1},a_{i}$ and their positional
relationship.

An important class of the random sequences is the \textit{binary}
chains where each symbol~$a_i$ can take on only two values, say,
$0$~and~$1$.

A very important subclass of the Markov chains is the
\textit{additive} binary ones. The conditional probability functions
for these chains are described by the following formula:
\begin{equation}\label{adm}
P_N(a_{i}=1|a_{i-N},\ldots,a_{i-2},a_{i-1}) =\bar{a}+\sum_{r=1}^{N}
F(r)(a_{i-r}-\bar{a}).
\end{equation}
Here $F(r)$,~$r=1,\ldots,N$, is the \textit{memory function} and
$\bar{a}$ is the average number of unities in the sequence (see,
e.g., Ref.~\cite{mel}).

A \textit{permutative} binary $N$-step Markov chain is determined by
the conditional probability that is independent of the order of
symbols within the memory length $N$ and depends on the number of
unities among them only.

We define an \textit{isotropic} chain as a sequence for which the
probability of an arbitrary set of $L$ sequential symbols occurring,
referred to as the \textit{$L$-word}, does not depend on the
direction of ''reading'' the symbols. For example, the probabilities
of $5$-words $01100$ and $00110$ occurring are equal.

In different mathematical and physical problems, we are confronted
with the sequences for which the conditional probability of
symbol~$a_i$ to have a certain value depends on values of~$N$
\textbf{previous} and~$N$ \textbf{next} symbols only,
\[P(a_i=a|\ldots,a_{i-2},a_{i-1},a_{i+1},a_{i+2},\ldots)=
\quad\quad\phantom{stas}
\]
\begin{equation}\label{def_2s}
\phantom{stas}\quad\quad
=P_{2,N}(a_i=a|a_{i-N},\ldots,a_{i-1},a_{i+1},\ldots,a_{i+N}).
\end{equation}
We refer to these chains as~\textit{$N$-two-sided} sequences. One
can define \emph{additive} binary two-sided chains similarly to the
Markov ones,
\begin{equation}\label{def_2sa}
P_{2,N}(a_{i}=1|a_{i-N},\ldots,a_{i-1},a_{i+1},\ldots,a_{i+N})=
\bar{a}+\sum_{\substack{r=-N\\r\neq 0}}^{N} G(r)(a_{i+r}-\bar{a}).
\end{equation}
Here $G(r)$, $r=\pm 1,\ldots,\pm N$, is the \emph{memory function of
the two-sided chain}. For the same reason as above, see
Eq.~(\ref{def_mark}), this chain is homogeneous.

In paper~\cite{equiv}, the equivalence of the two-sided and the
Markov chains was proved. We have derived the equation describing
the correspondence between their conditional probabilities,
\begin{equation}\label{dsv}
P_{2,N}(a_i=1|T_i^-,T_i^+)=\displaystyle\frac{
\prod\limits_{{r=0\atop a_i=1}}^N P_N(a_{i+r}|T^-_{i+r})}
{\prod\limits_{{r=0\atop a_i=1}}^N
P_N(a_{i+r}|T^-_{i+r})+\prod\limits_{{r=0\atop a_i=0}}^N
P_N(a_{i+r}|T^-_{i+r})}.
\end{equation}
Here $T^-_{j}=(a_{j-N},\ldots,a_{j-1})$ and
$T^+_{j}=(a_{j+1},\ldots,a_{j+N})$ are the previous and the
following $N$-words with respect to symbol~$a_j$.

Thus, any Markov chain can be characterized not only by the
conditional probability $P_N(a_{i}=a|T_i^-)$, but by the two-sided
conditional probability function, $P_{2,N}(a_i=a|T_i^-,T_i^+)$, as
well.
%%%%%%%%%%%%%%%%%%%%%%%%%%%%%%%%%%%%%%%%%%%%%%%%%%%%%%%%%%%%
\section{Isotropy of the Markov chains}
%%%%%%%%%%%%%%%%%%%%%%%%%%%%%%%%%%%%%%%%%%%%%%%%%%%%%%%%%%%%
\subsection{Two-sided chains with symmetric probability function}

The definition of isotropic $N$-step Markov chain given in
Sec.~\ref{def} is equivalent to the following statement: the Markov
chain is isotropic if its two-sided conditional probability function
is symmetrical,
\begin{equation}\label{imc}
P_{2,N}(a_i=a|{T}^{-}_{\!\rightarrow},T^{+}_{\!\rightarrow})
=P_{2,N}(a_i=a|T^{+}_{\!\leftarrow},T^{-}_{\!\leftarrow}).
\end{equation}
Here $T^-$ and $T^+$ are the previous and the following $N$-words
with respect to symbol $a$. The subscript ``$_{\!\rightarrow}$''
indicates that word~$T$ is read in the direct order, from left to
right, and the subscript ``$_{\!\leftarrow}$'' shows that word~$T$
is read in the inverse order. For example, if the binary
$3$-two-sided chain is isotropic, then
\[P_{2,3}(a_i=1|011,001)=P_{2,3}(a_i=1|100,110).\]
Equation~(\ref{imc}) can serve as the second definition of the
isotropy for the Markov chains.

Below we prove the equivalence of these two definitions of the
isotropy. This equivalence is the main result of our paper.

Suppose that the first definition is valid and, therefore, the
probability $P({T}^{-}_{\!\rightarrow},a_j=a,T^{+}_{\!\rightarrow})$
of an arbitrary $(2N+1)$-word occurring in the Markov chain is
independent of the direction in which it is read. Then, the formula
for two-sided probabilities can be rewritten as
\[P_{2,N}(a_j=a|{T}^{-}_{\!\rightarrow},T^{+}_{\!\rightarrow})=
\frac{P({T}^{-}_{\!\rightarrow},a_j=a,T^{+}_{\!\rightarrow})}{
\sum\limits_b
P({T}^{-}_{\!\rightarrow},a_j=b,T^{+}_{\!\rightarrow})}=
\frac{P(T^{+}_{\!\leftarrow},a_j=a,T^{-}_{\!\leftarrow})}{
\sum\limits_b P(T^{+}_{\!\leftarrow},a_j=b,T^{-}_{\!\leftarrow})}
=P_{2,N}(a_j=a|T^{+}_{\!\leftarrow},T^{-}_{\!\leftarrow}).\] Thus,
the two-sided conditional probability function of the Markov chain
is symmetrical and the chain is isotropic in accordance with the
second definition.

Now, let us suppose that the second definition of the isotropy is
valid, i.e., the two-sided probability of the Markov chain is
symmetrical.

%%%%%%%%%%%%%%%

Then, this chain, being read in the direct and inverse orders, is
statistically identical.

In other words, the probability of symbol~$a_i=a$ occurring under
condition that \textbf{previous} $N$-word~$T_{\!\rightarrow}$ is
fixed is equal to that of the same symbol occurring under the
condition that the \textbf{following} $N$-word~$T_{\!\leftarrow}$ is
fixed. It means that the original chain and its copy written in the
inverse order have the same Markov conditional probabilities.
However, the probabilities of words of arbitrary length~$L$ are
determined completely by the conditional probabilities. Hence, these
 probabilities are the same for both chains.

Thus, the probability of an arbitrary word of length~$L$ occurring
does not depend on the direction in which it is read.

%%%%%%%%%%%%

According to Eq.~(\ref{dsv}) this chain, being read in the direct
and inverse orders, has the same Markov conditional probabilities.
However, the probabilities of words of arbitrary length~$L$ are
fully governed by the conditional probabilities. Hence, this chain
is isotropic according the first definition.

%%%%%%%%%%%%%

In the general case, the Markov chains are anisotropic.
Nevertheless, our analysis of Eq.~(\ref{dsv}) shows that all 1-step
and 2-step additive Markov chains are isotropic. All
\textbf{non-biased} 3-step additive Markov chains are also
isotropic. The additive chains can be anisotropic for $N\geqslant
3$. The $3$-step biased Markov chains with conditional probability
\begin{equation}\label{four}
P(a_i=1|a_{i-3}, a_{i-2}, a_{i-1})=\bar{a}+\sum\limits_{r=1}^3 f_r
(a_{i-r}-\bar{a})
\end{equation}
and $\bar{a}\not=1/2$ are isotropic in three exceptional cases only:
\begin{enumerate}
   \item $f_1=f_2$.  This condition is fulfilled, e.g., for the
chains with the step-wise memory function.
   \item $f_1+f_2+f_3=1$. It is the degenerated case. This
memory function determines the Markov chain consisting completely of
unities or zeroes, since $P(a_i=1|111)=P(a_i=0|000)=1$.
   \item $f_3=0$. Actually, it is the $2$-step additive Markov chain.
\end{enumerate}

All additive Markov chains with small memory functions, $F(r)
\propto \varepsilon \ll 1$, are isotropic in the main approximation
with respect to $\varepsilon$. This fact will be clarified in
Subsec.~\ref{small1}.

It should be noted that the two-sided chain with asymmetrical
conditional probability function can be considered as the Markov
chain being read in the direct order and as another Markov chain
being read in the inverse order. The conditional probability
functions of these chains are different. From the foregoing two
conclusions can be made:

1. There are, at least, two different asymmetrical $N$-steps Markov
chains with equal correlation functions,
\begin{equation}\label{Def_K}
K(r)=\overline{(a_i-\bar{a})(a_{i+r}-\bar{a})}.
\end{equation}

2. The additive anisotropic Markov chain being read in the inverse
order is the non-additive Markov chain. Otherwise we would have two
additive Markov chains with different memory functions having the
same correlation function. But as shown in Ref.~\cite{AllMemStepCor}
it is not feasible.

%%%%%%%%%%%%%%%%%%%%%%%%%%%%%%%%%%%%%%%%%%%%%%%%%%%%%%%%%%%%
\subsection{Permutative Markov chains}
%%%%%%%%%%%%%%%%%%%%%%%%%%%%%%%%%%%%%%%%%%%%%%%%%%%%%%%%%%%%
\subsubsection{Isotropy of permutative Markov chains}

Here we prove that the permutative binary $N$-step Markov chain is
isotropic. Let us consider two chains, $\mathrm{M}$ and
$\mathrm{M}'$, where $\mathrm{M}$ is the given Markov chain,
$\{a_i\}$, and $\mathrm{M}'$ is the chain written in the inverse
order, $\{a'_i\}$, $a'_i=a_{-i}$. We refer to
$\overrightarrow{P}_k$, or
$\overrightarrow{P}(a_i=1|T(k))=\overrightarrow{P}(1|T(k))$, as the
probability of symbol $a_i$ in chain $\mathrm{M}$ to be equal to
unity given the previous $N$-word $T(k)$ contains $k$ unities. The
probability of the $N$-word occurring satisfies the set of linear
equations,
\begin{equation}\label{for_word}
\left\{\begin{array}{l} P(a_1,a_2,\ldots, a_N)=
\sum\limits_{a_0}\overrightarrow{P}(a_N|a_0,a_1,\ldots,
a_{N-1})P(a_0,a_1,\ldots,
a_{N-1}),\\
\sum\limits_{a_1}\ldots \sum\limits_{a_N} P(a_1,a_2,\ldots,
a_N)=1.
\end{array}
\right.
\end{equation}
The probability $P(a_1,a_2,\ldots, a_N)$ of $N$-word occurring is
determined uniquely by Eq.~\eqref{for_word}. The solution of this
set of equations depends on the total number $k=a_1+a_2+\ldots+ a_N$
of the unities in the $N$-word $(a_1,a_2,\ldots, a_N)$ and can be
presented in the form $P(a_1,a_2,\ldots, a_N)=P_k$. One can easily
find the following expression for $P_k$,
\[P_k=P_0\prod_{r=1}^{k}\dfrac{\overrightarrow{P}_{r-1}}
{1-\overrightarrow{P}_r}.\]

We also refer to $\overleftarrow{P}(1|T)$ as the probability of
symbol $a'_i$ to be equal to unity under condition that the previous
$N$-word in chain $\mathrm{M'}$, $T$, is fixed. Now let us prove
that the conditional probability functions of chains $\mathrm{M}$
and $\mathrm{M}'$ are equal to each other. In other words,
$\overleftarrow{P}(1|T)$ equals to $\overrightarrow{P}_k$ for
arbitrary $N$-word $T$ containing exactly $k$ unities
($k=1,\ldots,N$) and is not dependent on the order of symbols in
this word. With this purpose in mind one needs to prove that
\[\overleftarrow{P}(1|[k],0)=\overleftarrow{P}(1|[k-1],1)=
\overrightarrow{P}_k.\] Here $[j]$ means $(N-1)$-word containing
exactly $j$ unities, so that $(1,[k-1])$ and $([k-1],1)$ are the
$N$-words containing $k$ unities. Using the definition of the
conditional probability we find:
\[\overleftarrow{P}(1|[k-1],1)=
\overleftarrow{P}(1|[k-1],1) \frac{P_k} {P_k}=
\frac{P(1,[k-1],1)}{P_k}=
\overrightarrow{P}(1|1,[k-1])=
\overrightarrow{P}_k,\]

\[\overleftarrow{P}(1|[k],0)=
 \frac{P(1,[k],0)} {P_k}=\big(1- \overrightarrow{P}_{k+1}\big) \frac{P_{k+1}}
{P_k}.\]

From set~(\ref{for_word}), one gets the relation
\[P_{k+1}=\overrightarrow{P}_{k+1}{P_{k+1}}+
\overrightarrow{P}_{k}{P_{k}}.\] So, we have proved that
\[\overleftarrow{P}(1|[k-1],1)
=\overleftarrow{P}(1|[k],0)=\overrightarrow{P}_k.\] Thus we can
refer to $\overleftarrow{P}_k$ as the probability of symbol in chain
$\mathrm{M}'$ to be equal to unity under condition that the previous
$N$-word containing exactly $k$ unities is fixed, and the
conditional probability functions for the chains $\mathrm{M}$ and
$\mathrm{M}'$ are equal to each other,
\begin{equation} \label{Pk=Pk}
\overrightarrow{P}_k=\overleftarrow{P}_k,\,k\geqslant1.
\end{equation}
Hence, \begin{equation} \label{P0=P0}
\overrightarrow{P}_0=\overleftarrow{P}_0
\end{equation} since
\[\sum_T {\overrightarrow{P}}(1|T)P(T)=
\sum_T \overleftarrow{{P}}(1|T)P(T)=1.\]

Equations~\eqref{Pk=Pk} and~\eqref{P0=P0} do imply that the
permutative binary $N$-step Markov chains are isotropic.
%%%%%%%%%%%%%%%%%%%%%%%%%%%%%%%%%%%%%%%%%%%%%%%%%%%%%%%%%%%%
\subsubsection{Two-sided probability functions of
permutative Markov chains}

As was mentioned above (see Eq.~\eqref{dsv}), every Markov chain can
be regarded as two-sided one. Below we examine the properties of the
two-sided conditional probability function of the Markov chains with
one-sided conditional probability functions possessing the property
of permutability. An essential point is that this property does not
provide the permutability of the two-sided conditional probability
function. To demonstrate this fact we will show that in the general
case the two-sided conditional probability function changes its
value when two neighboring symbols $1$ and $0$ are transposed.

Let us prove this statement by contradiction and suppose that the
two-sided conditional probability function $P(a_i=1|T^-_i,T^+_i)$
takes on the same value for two variants of the word $T^-_i$. These
variants of the word $T^-_i$ only differ in the values of the
neighboring symbols $a_j$ and $a_{j+1}$: $a_j=1, a_{j+1}=0$ in the
first variant and $a_j=0, a_{j+1}=1$ in the second one. Consider the
structure of Eq.~\eqref{dsv}. Taking into account the permutability
of the Markov conditional probability function one can see that $N$
factors in all products in Eq.~\eqref{dsv} coincide for two symbols
sets under study. In the general case, only one factor,
$P_N(a_{j+N+1}|T^-_{j+N+1})$, changes its value. If $a_{j+N+1}=1$,
the coincidence of the values of $P(a_i=1|T^-_i,T^+_i)$ for two sets
of symbols under the consideration yields the following relation
between the values of one-sided conditional probability function:
\begin{equation}
\overrightarrow{P}^2_{k}=\overrightarrow{P}_{k+1}
\overrightarrow{P}_{k-1}
\end{equation}
for $k=a_{j+2}+a_{j+3}\ldots+a_{j+N}$, $1 \leqslant k \leqslant
N-1$. In the opposite case, at $a_{j+N+1}=0$, the similar
requirement is
\begin{equation}
(1-\overrightarrow{P}_{k})^2=(1-\overrightarrow{P}_{k+1})(1-
\overrightarrow{P}_{k-1}).
\end{equation}
These two relations are compatible for the non-correlated chain only
where $\overrightarrow{P}_{k}$ is $k$-independent. The exception is
the case of one-step binary Markov chain that always has permutative
two-sided conditional probability function.

Thus, any correlated multi-step permutative Markov chain possesses
the non-permutative two-sided probability function.

%%%%%%%%%%%%%%%%%%%%%%%%%%%%%%%%%%%%%%%%%%%%%%%%%%%%%%%%%%%%
\subsection{Additive weakly correlated Markov chains}

The third class of the isotropic sequences represents the additive
weakly correlated Markov chains. For these chains, we suppose that
the memory function is small,
\begin{equation}\label{small}
\sum\limits_{r=1}^N|F(r)|\ll 1.
\end{equation}
Their asymptotical isotropy is proved in Subsection~\ref{small1}.

Every function of $N$ variables, $f(a_1,\dots,a_N)$, satisfying the
evident restriction, $0\leqslant f(a_1,\dots,a_N) \leqslant 1$, can
be thought of the conditional probability function
$P(a_i=1|a_{i-N},\ldots,a_{i-2},a_{i-1})$ of some Markov chain. Yet
not every function of $2N$ variables, even if restricted by the
similar condition, is the conditional probability function of some
two-sided chain. It follows from Eq.~(\ref{dsv}), that an arbitrary
binary Markov chain is determined by $2^N$ parameters, i.e. the
number of all possible sets of arguments in the conditional
probability function. Hence, a two-sided chain equivalent to this
Markov one, is also determined by $2^N$ parameters. Nevertheless,
the two-sided conditional probability function
$P(a_i=1|T^-_i,T^+_i)$ is formally governed by $2^{2N}$ parameters,
i.e. the number of all possible \textbf{previous}, $T^-_i$, and
\textbf{following}, $T^+_i$, $N$-words. So, not every function of
$2N$ arguments can play a role of some two-sided conditional
probability function. The example is an additive two-sided chain
with small memory function $G(r)$ in Eq.~(\ref{def_2sa}). In this
case, $G(r)$ has to be asymptotically even. This fact is proven in
the Subsection~\ref{small2}.

%%%%%%%%%%%%%%%%%%%%%%%%%%%%%%%%%%%%%%%%%%%%%%%%%%%%%%%%%%%%

\subsubsection{Isotropy of additive weakly correlated Markov chains\label{small1}}

In order to find the conditional probability function,
$P(a_i=1|T^-_{i},T^+_{i})$, of the weakly correlated additive Markov
chain one has to substitute the probability Eq.~\eqref{adm} into
Eq.~\eqref{dsv}, and retain the terms of the zeroth and first orders
in $F(r)$. The obtained two-sided conditional probability function
takes the form of  Eq.~(\ref{def_2sa}) with even memory function:
$G(r)=G(-r)=F(r)$. So, the additive weakly correlated Markov chains
are asymptotically isotropic.

%%%%%%%%%%%%%%%%%%%%%%%%%%%%%%%%%%%%%%%%%%%%%%%%%%%%%%%%%%%%

\subsubsection{Restriction on the class of
the memory functions of weakly correlated additive two-sided
chains}\label{small2}

In this subsection, we show that the weakly correlated additive
two-sided chain is asymptotically isotropic, i.e. the two-sided
memory function $G(r)$ is necessarily even. To this end we consider
arbitrary two-sided additive chain and find its one-sided
conditional probability function, $P(a_i=1|T^-_i)$. We will prove
that this function is reduced to the additive form, Eq.~(\ref{adm}),
with the memory function $F(r)=G(-r)$, and, therefore, the chain
under consideration is asymptotically isotropic.

Let us examine the additive two-sided chain (not obligatory
isotropic) and find its one-sided conditional probability function.
In the general case the problem is reduced to solving the set of
$2^N$ non-linear equations, Eq.~\eqref{dsv}, written for different
sequences of symbols in word $T_i^-$. We consider weakly correlated
additive two-sided chain subjected to the restriction
$\sum\limits_{r=1}^N \big(|G(-r)|+|G(r)| \big)\ll 1$. Its one-sided
conditional probability function can be presented in more convenient
form:
\begin{equation}\label{vid}
P(a_i|T^-_{i})= 1-a_i+(2a_i-1)\big(\bar{a}+\varphi(T^-_{i})\big),
\end{equation}
with function $\varphi(T^-_{i})$ to be determined. The evident
equation $P(a_i=1|T^-_{i})=1-P(a_i=0|T^-_{i})$ is fulfilled for
Eq.~\eqref{vid}. If $G(r)$ tends to zero, the probability
$P(a_i=1|T^-_{i})$ goes to $\bar{a}$ and function $\varphi$ tends to
zero: $\varphi(T^-_{i})\to 0$. Now, substituting the conditional
probability function in the form of Eq.~\eqref{vid} into
Eq.~\eqref{dsv} and retaining only terms of the zeroth and first
orders in $\varphi(T^-_{i})$, we obtain the approximate expression
for the two-sided conditional probability function:
\begin{equation}
P(a_i=1|T^-_{i},T^+_{i})\simeq
\bar{a}\Big(1+(1-\bar{a})\Big(\sum\limits_{\substack{r=0\\a_i=1}}^N
\psi(a_{i+r})\varphi(T^-_{i+r})
-\sum\limits_{\substack{r=0\\a_i=0}}^N
\psi(a_{i+r})\varphi(T^-_{i+r})\Big)\Big).\label{dvser}
\end{equation}
Here we introduce a new function $\psi$,
\begin{equation}
\psi(a_i)=\dfrac{a_i-\bar{a}}{\bar{a}(1-\bar{a})}=\begin{cases}
1/(\bar{a}-1), & a_i=0\\
{1}/{\bar{a}},& a_i=1. \end{cases}
\end{equation}
The two-sided conditional probability function of the chain under
consideration is given by Eq.~\eqref{def_2sa}. So, finally, we
obtain
\[\sum\limits_{\substack{r=0\\a_i=1}}^N
\psi(a_{i+r})\varphi(T^-_{i+r})-\sum\limits_{\substack{r=0\\a_i=0}}^N
\psi(a_{i+r})\varphi(T^-_{i+r})\]\begin{equation}
\simeq\frac{1}{\bar{a}(1-\bar{a})}\sum_{r=1}^N\big(G(r)(a_{i+r}-\bar{a})
+G(-r)(a_{i-r}-\bar{a})\big). \label{usl2}
\end{equation}
Calculating the difference between two expressions presented by
Eq.~\eqref{usl2} written for $a_{i+N}=1$ and $a_{i+N}=0$, we get
\begin{equation}\label{Ndif}
\left. \varphi(T^-_{i+N})\right|_{a_i=1}-\left.
\varphi(T^-_{i+N})\right|_{a_i=0}\simeq G(N).
\end{equation}
Substitution of Eq.~\eqref{Ndif} in Eq.~\eqref{usl2} yields,
\begin{equation}\notag
\sum\limits_{\substack{r=0\\a_i=1}}^{N-1}
\psi(a_{i+r})\varphi(T^-_{i+r})-
\sum\limits_{\substack{r=0\\a_i=0}}^{N-1}
\psi(a_{i+r})\varphi(T^-_{i+r})
\end{equation}
\begin{equation}\label{usl2-1}
\simeq \frac{1}{\bar{a}(1-\bar{a})}
\Big(\sum_{r=1}^{N-1}\big(G(r)(a_{i+r}-\bar{a})
+G(-r)(a_{i-r}-\bar{a})\big)+ G(-N)(a_{i-N}-\bar{a})\Big).
\end{equation}
Now we repeat this procedure $N-1$ times. At the first repeat we
calculate the difference between two expressions~\eqref{usl2-1}
written for $a_{i+N-1}=1$ and $a_{i+N-1}=0$ and substitute the
obtained result in Eq.~\eqref{usl2-1}, and so on. At the last repeat
we obtain,
\begin{equation}\label{last}
\varphi(T^-_{i})\simeq \sum_{r=1}^NG(-r)(a_{i-r}-\bar{a}).
\end{equation}
So the one-sided conditional probability function is
\begin{equation}\label{ans}
P(a_i=1|T^-_{i})\simeq\bar{a}+ \sum_{r=1}^NG(-r)(a_{i-r}-\bar{a}).
\end{equation}
As it follows from previous Subsection, Markov chains with such
conditional probability functions are isotropic. Thus, we have found
that the additive weakly correlated two-sided chain is
asymptotically isotropic. In other words, the memory function of
additive weakly correlated chain can be only even, $G(-r)=G(r)$.
%%%%%%%%%%%%%%%%%%%%%%%%%%%%%%%%%%%%%%%%%%%%%%%%%%%%%%%%%%%%
\section{Conclusion}

Thus, using the equivalence of the Markov and two-sided chains, we
studied the important property of the Markov chains, their isotropy.
The results of this study are shown in the scheme. Here,
$A-\!\!st\!\!\!\rightarrow \! B$ means, that the chains from class
$A$, restricted by the statement $st$, are the members of class $B$.
The most evident fact is that the Markov chains, that possess
symmetric two-sided conditional probability function, are isotropic.
Another important class of the isotropic Markov chains are the
sequences with the permutative conditional probability functions.
One of the examples of such chains are the additive Markov chains
with the step-wise memory functions, examined in details in
Refs.~\cite{uyakm, AllMemStepCor}. The additive weakly correlated
chains are also isotropic. Such chains play a key role in the
non-extensive thermodynamics of Ising chains of classical spins with
long-range interaction, as well as in the literary texts and
sequences of nucleotides in DNA molecules.

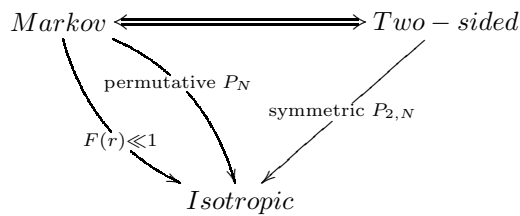
\[
\xymatrix{
 Markov \ar@{<=>}[rr]
\ar@{->}@/^1.3pc/[ddr]|{\mbox{\scriptsize permutative } P_N}
\ar@{->}@/_1.3pc/[ddr]|{F(r)\ll 1} & & Two-sided
 \ar@{->}[ddl]|{\mbox{\scriptsize symmetric } P_{2,N}}\\ &&\\
 & Isotropic & }
\]


\begin{references}

\bibitem{bul} U. Balucani, M. H. Lee, V. Tognetti, Phys.
Rep. \textbf{373}, 409 (2003).

\bibitem{sok} I. M. Sokolov, \prl \textbf{90}, 080601 (2003).

\bibitem{bun} A. Bunde, S. Havlin, E. Koscienly-Bunde,
H.-J. Schellenhuber, Physica A \textbf{302}, 255 (2001).

\bibitem{yan} H. N. Yang, Y.-P. Zhao, A. Chan, T.-M. Lu, and
G. C. Wang, \prb \textbf{56}, 4224 (1997).

\bibitem{maj} S. N. Majumdar, A. J. Bray, S. J. Cornell, and
C. Sire, \prl \textbf{77}, 3704 (1996).

\bibitem{halvin} S. Halvin, R. Selinger, M. Schwartz,
H. E. Stanley, and A. Bunde, \prl \textbf{61}, 1438 (1988).

\bibitem{vossDNA} R. F. Voss, \prl \textbf{68}, 3805 (1992).

\bibitem{stan}  H.~E.~Stanley \textit{et. al.}, Physica A
\textbf{224},302 (1996).

\bibitem{buld} S. V. Buldyrev, A. L. Goldberger, S. Havlin, R. N.
Mantegna, M. E. Matsa, C.-K. Peng, M. Simons, H. E. Stanley, \pre
\textbf{51}, 5084 (1995).

\bibitem{prov}  A. Provata and Y. Almirantis, Physica A \textbf{247},
482 (1997).

\bibitem{yul} R. M. Yulmetyev, N. Emelyanova, P. H\"{a}nggi, and
F. Gafarov, A. Prohorov, Phycica A \textbf{316}, 671 (2002).

\bibitem{hao} B. Hao, J. Qi, Mod. Phys. Lett., \textbf{17}, 1 (2003).

\bibitem{mant}  R. N. Mantegna, H.~E. Stanley, Nature (London)
\textbf{376}, 46 (1995).

\bibitem{zhang}  Y. C. Zhang, Europhys. News, \textbf{29}, 51 (1998).

\bibitem{schen}  A. Schenkel, J. Zhang, and Y. C.~Zhang,
Fractals \textbf{1}, 47 (1993).

\bibitem{kant}  I. Kanter and D. A. Kessler, \prl \textbf{74}, 4559
(1995).

\bibitem{kokol} P. Kokol, V. Podgorelec, Complexity International,
\textbf{7}, 1 (2000).

\bibitem{ebeling} W. Ebeling, A. Neiman, T. Poschel,
arXiv:cond-mat/0204076.

\bibitem{uyakm} O. V. Usatenko, V. A. Yampol'skii, K. E. Kechedzhy
and  S. S. Mel'nyk, Phys. Rev. E \textbf{68}, 06117 (2003).

\bibitem{uya}  O. V. Usatenko and V. A. Yampol'skii, \prl \textbf{90},
110601 (2003).

\bibitem{mel} S. S. Melnyk, O. V. Usatenko, and V. A. Yampol'skii,
Physica A, \textbf{361}, 405 (2006); arXiv:physics/0412169.

\bibitem{AllMemStepCor} S. S. Melnyk, O. V. Usatenko,
V. A. Yampol'skii, S. S. Apostolov, and Z. A. Mayzelis,
arXiv:physics/0306170.
%; to be published in Journ. of Phys. A.

\bibitem{tsal} C. Tsalis, J.\ Stat.\ Phis. \textbf{52}, 479
(1988).

\bibitem{abe} \textit{Nonextensive Statistical Mechanics and Its
Applications}, eds. S. Abe and Yu. Okamoto (Springer, Berlin, 2001).

\bibitem{den} S. Denisov, Phys.\ Lett.\ A, \textbf{235}, 447
(1997).

\bibitem{equiv} S. S. Apostolov, and Z.A. Mayzelis, O. V. Usatenko,
and V. A. Yampol'skii, arXiv:physics/0306170.

\end{references}
\end{document}